\documentclass[preprints,article,accept,pdftex,moreauthors]{Definitions/mdpi}

\firstpage{1} 
\pubvolume{1}
\issuenum{1}
\articlenumber{0}
\pubyear{2025}
\copyrightyear{2025}
 
\datereceived{ } 
\daterevised{ }  
\dateaccepted{ } 
\datepublished{ }

\hreflink{https://doi.org/}

\usepackage{makecell}

\Title{Joint Data Hiding and Partial Encryption of Compressive Sensed Streams}

\TitleCitation{Joint Data Hiding and Partial Encryption of Compressive Sensed Streams}

\Author{Cristina-Elena Popa$^{1}$, Cristian Damian $^{1}$* and Daniela Coltuc $^{1,}$}

\AuthorNames{Cristina-Elena Popa, Cristian Damian and Daniela Coltuc}

\isAPAStyle{%
       \AuthorCitation{Popa, C.E., Damian, C.C., \& Coltuc, D.}
         }{%
        \isChicagoStyle{%
        \AuthorCitation{Cristina-Elena Popa, Cristian-Constantin Damian, and Daniela Coltuc.}
        }{
        \AuthorCitation{Popa, C.E.; Damian, C.C.; Coltuc, D.}
        }
}

\address{%
$^{1}$ \quad National University for Science and Technology POLITEHNICA Bucharest, 060042 Romania\\}

\corres{Correspondence: constantin.damian91@upb.ro}

\abstract{The paper proposes a method to secure the Compressive Sensing (CS) streams. It consists in protecting part of the measurements by a secret key and inserting the code into the rest. The secret key is generated via a cryptographically secure pseudo-random number generator (CSPRNG) and XORed with the measurements to be inserted. For insertion, we use a reversible data hiding (RDH) scheme, which is a prediction error expansion algorithm, modified to match the statistics of CS measurements. The reconstruction from the embedded stream conducts to visibly distorted images. The image distortion is controlled by the number of embedded levels. In our tests, the embedding on 10 levels \textcolor{black}{ results in $\approx 18 dB $ distortion} for images of 256x256 pixels reconstructed with the Fast Iterative Shrinkage-Thresholding Algorithm (FISTA) \cite{Beck2009Fast}. A particularity of the presented method is on-the-fly insertion that makes it appropriate for the sequential acquisition of measurements by a Single Pixel Camera. On-the-fly insertion avoids the buffering of CS measurements for a subsequent standard encryption and generation of a thumbnail image.}


\begin{document}

\section{Introduction}

The single pixel camera (SPC) concept is an application of the Compressive Sensing (CS) paradigm in which signal compression is done at acquisition time by exposing a 2D scene to a sequence of masks and recording the light reflected by each of them using a single sensor. The captured scene can be reconstructed based on these SPC measurements with the use of various reconstruction algorithms.
In order for the reconstruction to be possible, the original scene must be sparse in some domain and the columns of the sensing matrix must be incoherent with the sparsity basis.

Such cameras are an alternative to widely used technologies such as charge-couple devices (CCD) or metal-oxide-semiconductor (CMOS) that are only available for a small section of the electromagnetic spectrum. Applications for SPC have been developed in multiple fields, such as fluorescence \cite{Yang2018FluorescenceMicroscope} and infrared \cite{Radwell2014InfraredMicroscope} microscopy, methane gas leak detection \cite{Graham2017MehaneLeaks} and three dimensional imaging \cite{Zhang20163DImaging}.

Today, most applications include communication on public channels, where the data are exposed at  theft or attacks.
The common practice to protect the data is standard encryption after the compression stage.  
\textcolor{black}{Another solution is the partial encryption where, in the case of transform-based compression of images, only the sign or part of significant coefficients are encrypted. One difference from the standard encryption is the  resulted image, which is noise-like in standard encryption and with discernable content in partial encryption. }  
The main drawback of the partial encryption is the weakness to error concealment attacks (ECA). It was shown that the encrypted coefficients can be estimated by using semantic or statistical rationales to obtain an image closer to the original.

\textcolor{black}{Data hiding} is the generic name for  the set of techniques  aiming to insert information in a multimedia carrier. 

\textcolor{black}{The reversible data hiding (RDH)} is a particular class of methods allowing \textcolor{black}{perfect recovery of the carrier after information} extraction.

\textcolor{black}{Data hiding is commonly used for data integrity, covert communication, non-repudiation, copyright protection, and authentication, among other applications \cite{Megias2021}.}

\textcolor{black}{We propose an RDH method, specific to CS, that consists in protecting part of the CS measurements by a secret key and embedding them into the rest. The method can also be seen as an alternative to partial encryption, due to the result which consists of reconstructed images with visible distortion.}

\textcolor{black}{The embedded measurements are protected by XOR with a single-use secret key generated via a cryptographically secure pseudorandom number generator (CSPRNG) that passes the NIST SP 800-22 tests \cite{Almaraz2023CSPRNG}. The key’s length is larger than the encrypted plaintext. The secret key must be transmitted using a secure communication protocol that follows the guidelines in NIST SP 800-57 \cite{NIST800-57}, thus ensuring that only authorized parties can access it.}

The embedding is done on-the-fly, by a RDH method, which is a prediction error expansion algorithm modified such to fit CS measurements statistics. 

Opposite to the common practice in RDH, 
where the concern is to embed data without visible distortion of the carrier, in our method the embedding is done on a high number of bits such to cause strong visual distortion if an unauthorized user tries to reconstruct the image from the 
 
\textcolor{black}{modified} measurements. 

\textcolor{black}{The method is dedicated to the confidential information protection of CS measurements obtained sequentially, from an SPC. }Due to the on-the-fly insertion, it avoids the buffering of CS measurements for a subsequent standard encryption and generation of a thumbnail preview. 

The novelty of our method is:
\begin{itemize}
    \item  
    \textcolor{black}{Light encryption} by a secret key of part of CS measurements and use the rest as carrier. No selection for being embedded is done since in CS, the random projections are equipotent in image reconstruction. The image distortion is tuned by the percentage of embedded measurements. 
    \item On-the-fly embedding of the measurements. In a SPC scenario, the CS measurements are taken sequentially. To avoid the measurement buffering before transmission, the embedding is done as they are obtained.  
    \item A modified version of prediction error expansion algorithm. One particularity of the CS measurements is the statistical independence. Since there is no correlation between the neighboring measurements, the prediction error is uniformly distributed. To have a Gaussian distribution of the prediction error, we replaced the mean of neighbors with the mean of all measurements.      
\end{itemize}

The method is evaluated under the following aspects: 

\textcolor{black}{the capacity of the data hiding algorithm, the distortion introduced by embedding and the impact on data volume.}
An equation is derived for the insertion capacity and the experimental rate-distortion curve of the method is given.

The experiments were performed on synthetic and real data. A sky image was divided into 359 patches and for each patch we calculated a series of random random projections in order to simulate the acquisition. The idea behind this choice was the potential of SPC in sky exploration, where bands of the electromagnetic spectrum like the far infrared are of interest. For such bands, there are not large areas of sensors. Besides, the signal is very weak \cite{Coluccia2019}. 
The real data were measured under white light by using our setup in \cite{Petrovici2016Single}.

The rest of the paper is organized as follows: 
\textcolor{black}{Section II presents the two related domains i.e., data protection by a secret key and partial or selective encryption.}
Section III describes on-the-fly data hiding, explains the modified data hiding algorithm and discusses the impact on data representation.
Section IV is dedicated to the insertion capacity, an equation for capacity is derived and the mechanism of distortion control is introduced.
Section V analyses the sources of distortion in order to better understand how to control it.
Section VI discusses 
 
the method behavior at error concealment attacks.
Section VII is dedicated to experiments on the sky image and real data. Examples of image reconstruction are depicted and an experimental rate-distortion curve is plotted. 
Finally, some conclusions are drawn.

\section{Related Work}

CS theory naturally encounters the domain of cryptography.
 Many CS-based cryptosystems are built on the randomness of the measurement matrix, a sine qua non condition to obtain compressed samples by signal projection. As randomness means secrecy, the measuring matrix can be used as a secret key, making compression and encryption a single-step operation. 
 This considerably improves efficiency compared to the common approach, where the image is first compressed and then encrypted. 
Many solutions in this direction were proposed \cite{Escamilla2016, Zhou2014,Pomuna2016}.

A criticism of these methods has been the size of the secret key that hindered the distribution and transmission. This stimulated the development of new measuring matrices, generated with chaotic sequences \cite{Yu2010,Gilbert2010}.

Another limitation of this kind of methods is the one-time usage of the measurement matrix in order to preserve the secrecy property. Generating a new measuring matrix for each fresh signal is costly and impractical in a series of applications.
 This drawback has boosted a second category of methods, where the encryption is applied to the measurements, leaving the matrix public. 
 Techniques inspired from the encryption of Shanon/Nyquist samples \cite{Dufaux2008, Shi1998, Wang2010, Wen2002} like the permuting or substitution of measurements, the sign encoding were such solutions, to enumerate only a few.

The encryption of measurements conducts to noise-like reconstructions that draws the attention of attackers. 
To circumvent this problem, a class of methods called visually meaningful images encryption (VMIE) has been developed. By such methods, a plain image is pre-encrypted and then embedded into a host image, making the secret less perceptible \cite{Kanso2017,Singh2018,Chai2017,HWang2019}. 
Here, encryption meets the field of data hiding that encompasses methods aiming to hide information by embedding it in the less significant bits of an image. 
\textcolor{black}{An example of VMIE is in \cite{Huo2023}, where the authors first encrypt one or several images and then embed the code into a host image. The encryption is done by scrambling  image pixels, then the result is compressed by CS. The code is embedded into the wavelet coefficients of the host image, on multiple bit planes. Another example is in \cite{Huang2023}, where the plain image is compressed and encrypted to obtain the secret image, then zeros are inserted to extend it at the initial size. The result is wavelet transformed and embedded into the wavelet coefficients of a carrier image. The encryption is done by a public key generated by a chaotic system and reinforced by an RSA algorithm.} 

\textcolor{black}{There is a plethora of data hiding algorithms. A comprehensive survey can be found in \cite{Shi2016}. The emergent techniques of deep learning have been adopted also in the data hiding domain. Recent applications of deep learning for data hiding in digital signals are especially oriented towards image-based data hiding \cite{Wang2023}. Also, deep learning can augment CS with faster reconstruction algorithms or better quality of the reconstruction \cite{Machi2023}. However, when it comes to RDH performed in the CS samples, at our best knowledge, deep learning is not currently integrated.}

The partial or selective encryption is another related category of methods. It consists in the partial encryption of the CS measurements. 
It has roots in the partial encryption of transform-based compressed images, where only a part of the coefficients are encrypted \cite{Chenk2000, Van2004, Said2005, Taneja2011, Cheng2000}.  
Unlike the visually meaningful encryption,  the distortion introduced by partial encryption is visible but the image content is not faded out.
An example of partial encryption of compressed sensing measurements is in \cite{Wang2019} where only the measurements sign is encypted by XOR with a secret binary sequence. The method is completed by embedding side information in the ecrypted measurements. The side information could include sensor’s ID, sensor’s location, time of measuring etc. The embedding is done by using a reversible histogram shifting based hiding scheme.

\section{The Embedding of CS Measurements}
\label{Sec III: On-the-fly}

The algorithm we use is a reversible data hiding method that was proposed in \cite{Popa2020} and further discussed in \cite{Popa2021}. 
It is a prediction error expansion schema, as it was presented in \cite{Sachnev2009}, where the authors innovate at the level of predictor and insert data bit by bit into a 2D image.

If the prediction error sits between two thresholds $[T_n, T_p]$, it is expanded and an information is inserted. Otherwise, the value is shifted so that no overlapping occurs, and extraction is lossless.

Considering the image to be acquired $x \in \mathbb{R} ^{S\times 1}$ is sparse and the set of patterns that are projected
\textcolor{black}{$\phi \in \mathbb{R} ^{L\times S}$, }
 
the obtained measurements based on the CS are:

\begin{equation} y = \phi x
\label{eq_cs}
\end{equation}

The prediction error expansion algorithm has been updated to better fit this scenario: inserting the data on-the-fly in the measurements acquired sequentially by a SPC. 

\textcolor{black}{Firstly, the predicted value is a constant equal to the mean value of the measurements. This is done because, as opposed to a natural image where neighboring }pixels have correlated values and allow for a reliable prediction, consecutive SPC measurements are statistically independent of each other. This also allows us to try to insert data in all the positions in the stream, as none of them must remain unchanged in order to compute prediction errors. 

\textcolor{black}{
Secondly, the original algorithm uses two thresholds for data insertion and extraction, as information is embedded only when predicted errors fall between the thresholds.} This directly impacts the capacity and distortion. In the proposed schema, the two thresholds can be used for the fine control of the image distortion or simply ignored.

Lastly, our method allows multiple bits to be inserted as opposed to only one, leading to a higher possible distortion. This aspect is discussed in more detail in the following subsections.

\begin{figure}
    \centering
    \includegraphics[width=0.7\linewidth]{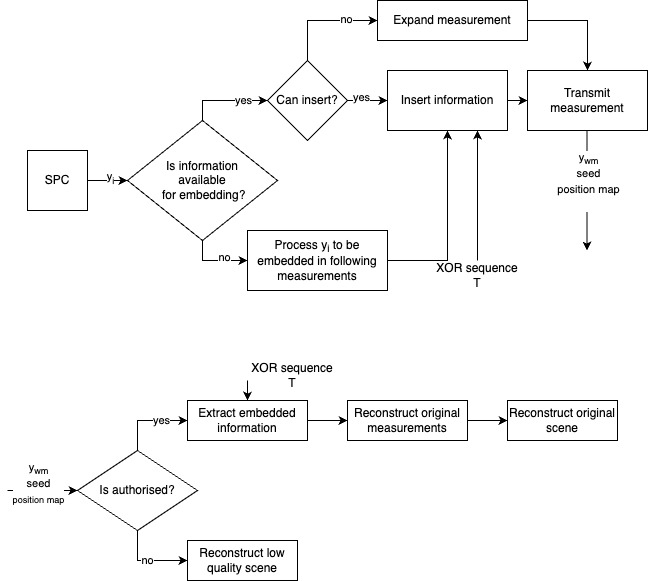}
    \caption{\textcolor{black}{Schema of the proposed scenario. (up) Data insertion: if information is available for embedding and the measurement is eligible, the modified measurement is computed and transmitted. If not, the measurement is shifted and transmitted. If information is no longer available for embedding, the current measurement is processed to be used for insertion. (down) Data extraction: if the user is authorised, the embedded information be extracted and the original measurements recovered. In the contrary case, the user is obliged to reconstruct the image using the reduced set of modified measurements}}
    \label{fig:wm-schema}
\end{figure}

Another important aspect is that the embedded data is composed of measurements that are processed and inserted on-the-fly at acquisition time (the flowchart is illustrated in Fig. \ref{fig:wm-schema}). This directly impacts the embedding capacity of the schema compared to the original one. 
\textcolor{black}{When selecting measurements to serve as embedded data there is no extra check done, the criteria used to decide if a value is eligible or not do not apply.}

The CS measurement matrix type is Hadamard with randomly permuted columns, a binary matrix that is easy to implement and use in CS scenarios \cite{Gan2008FastCI}. 
The construction of Hadamard matrix leads to a first row providing a measurement with a double maximum value compared to the others.
This is why the first row is not used in the \textcolor{black}{proposed}
 
scenario, being excluded altogether.

The columns are permuted using a random number generator with a known seed, so the results are reproducible by an end user. This seed is generated at acquisition time and is then transmitted with the measurements in order to make the reconstruction possible.

Two types of users can receive and use the \textcolor{black}{modified}
measurements: an authorised user that can \textcolor{black}{compute the original measurements}
and reconstruct the original image or an unauthorised user that \textcolor{black}{is only}
able to reconstruct a lower quality version of the image \textcolor{black}{using the modified measurements}.

In the case of an authorised user, the information received is: the \textcolor{black}{modified}
 
measurements, the seed used to scramble the Hadamard matrix, the position map of \textcolor{black}{embedded measurements and the secret key used for data protection.}

An unauthorised user only receives the \textcolor{black}{modified measurements and a reduced measurements matrix that can be used to reconstruct the distorted image. It is built by removing from the original matrix the columns corresponding to the indexes in the position map.}

\subsection{On-the-fly insertion}

\textcolor{black}{
Suppose that the SPC collects sensor measurements 
\textcolor{black}{$y \in \mathbb{R} ^{L\times 1}$} 
one by one and insertion is done on $n$ bits. In order to embed data in the acquired values, some of them are processed and used to mark the next measurements as follows:%
}

\begin{enumerate}
  \item \textcolor{black}{The measurement is first converted to binary (represented on 16 bits) and a XOR with the secret key is applied.}
  \item \textcolor{black}{The resulting 16 bits are split into chunks of length $n$}
  \item \textcolor{black}{Each chunk is converted back to decimal, resulting the values that will be inserted. If the division of the total number of bits to $n$ provides a remainder, the last bits are stored and appended to the next binary representation. } 
  \item \textcolor{black}{Each of the decimal values is inserted in the following measurement values based on the algorithm described in Section 3.2.}
\end{enumerate}

\textcolor{black}{The maximum for the inserted values} is:
\begin{equation}
bn_{max} = 2^n - 1
\label{bn_max}
\end{equation}

The insertion process is depicted in Fig. \ref{fig:wm-schema}. The SPC generates measurements $y_i$ one by one and each of them is processed at acquisition time. \textcolor{black}{The very first collected measurement is processed and the resulting $bn_i$ values are inserted in the next available measurements. For the following acquired measurements, if data is available for insertion and the measurement is eligible for marking, the data is embedded and the modified measurement value is transmitted.} If insertion is not possible, the measurement is shifted and transmitted. \textcolor{black}{If no data is available for insertion, the acquired measurement is processed to be embedded in the next coming ones.}

\begin{figure}
    \centering
    \includegraphics[width=0.8\linewidth]{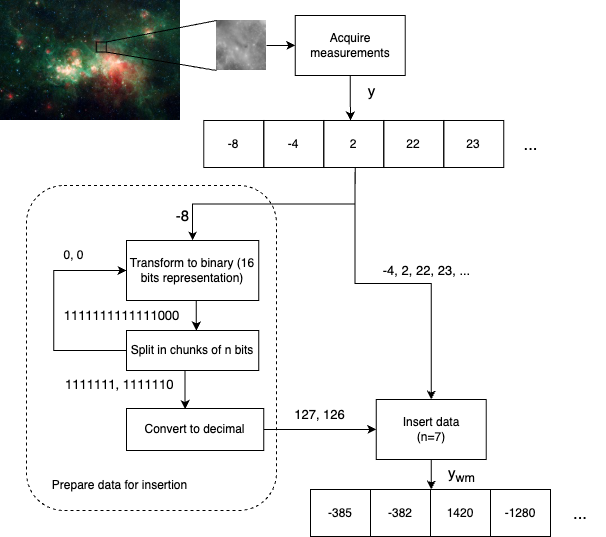}
    \caption{Embedding process example \textcolor{black}{on 7 bits}.}
    \label{fig:embedding_example}
\end{figure}

\textcolor{black}{A numerical example is depicted in Fig. \ref{fig:embedding_example}. The insertion is done on 7 bits, the threshold $T=10$ and the measurement to be inserted is $-8$. It is firstly converted in binary, to $[1,1,1,1,1,1,1,1,1,1,1,1,1,0,0,0]$. The 7 bits chunks are $[1,1,1,1,1,1,1]$ and $[1,1,1,1,1,1,0]$, a 2 bits remainder $[0,0]$ will serve to build up the next chunk. The carrying sequence consists of the following measurements [-4, 2, 22, 23]. The 2 chunks are inserted in the 1st and 2nd measurements that correspond to prediction errors less than T and larger than -T. The rest of the measurements are only shifted by 7 bits.}

With these steps, the following data is transmitted: the marked measurements $y_{wm}$, the seed value used to construct the measurement matrix and a location map representing the positions in $y$ that were extracted and \textcolor{black}{then embedded in following eligible measurements.}

An authorised user has access to extra information that allows the \textcolor{black}{original measurements to be computed}: the threshold value $T$ \textcolor{black}{(with $[-T, T]$ defining the expandable set of prediction errors)}, as well as the secret key used to perform the XOR operation.

Because part of the measurements are \textcolor{black}{used for embedding}
 
, the resulting $y_{wm}$ has a length of
\textcolor{black}{$L_{wm}$ that is less than $L$. }
 
This means that in order to be able to reconstruct a lower quality of the original image based on $y_{wm}$, an unauthorised user must use a truncated version of the measurement matrix. It can be constructed based on the random seed and location map as follows: generate a Hadamard matrix, permute the columns using the same random number generator algorithm and seed initially used and then remove the lines corresponding to the positions in the location map. 

Considering this, the extraction steps an authorised user would perform are as follows:

\begin{enumerate}
  \item \textcolor{black}{Go through the elements of $y_{wm}$ one by one to extract the embedded data and recover the set of original measurement values. This is possible since we use a RDH algorithm \cite{Popa2021}.}
  \item For each extracted value $bn_i$, convert it to binary, represent it on $n$ bits and perform a XOR operation \textcolor{black}{to obtain the original data.}
  \item Concatenate the resulting values and slice chunks of 16 bits. Each 16 bits value is then converted back to decimal, resulting the \textcolor{black}{set of measurement values embedded in the data insertion stage.}
  \item Use the position map to insert these in the correct locations\textcolor{black}{, finally obtaining the original values of $y$.}
\end{enumerate}

As the \textcolor{black}{deployed RDH} algorithm is reversible, the recovered values are the same as the original ones \textcolor{black}{except} the final one that could be truncated \cite{Popa2021}.  

\subsection{\textcolor{black}{Impact of data insertion on measurements dynamic range}}

As mentioned, given the particularities of CS measurements, we calculate the prediction error $d_i$ by considering the measurement mean $\overline{y}$.
Moreover, due to the construction of Hadamard matrix, the mean estimate is always zero:

\begin{equation}
d_i = y_i - \overline{y} =  y_i 
\label{d_i}
\end{equation}

\noindent It follows that the prediction error is the measurement itself and that the insertion is done directly into the measurements.

According to the algorithm in \cite{Sachnev2009}, the prediction error is expanded to $D_i$ in order to insert $bn_i$. \textcolor{black}{By supposing the threshold $T>0$ that defines the set of expandable measurements and the maximum possible inserted value $bn_{max}$ defined in eq. (\ref{bn_max}):}

\begin{equation}
   D_i =
    \begin{cases}
      2^n \cdot d_i + bn_i & \text{if $d_i \in [-T , T]$}\\
      d_i + bn_{max}T + bn_{max} & \text{if $d_i > T$}  
      \\
      d_i - bn_{max}T & \text{$d_i < -T$}  
    \end{cases}       
\label{D_i}
\end{equation}

\noindent In the insertion stage, the original $y_i$ value is replaced with the \textcolor{black}{marked} $y_{wm_i}$:

\begin{equation}
y_{wm_i} = D_i  + \overline{y} = D_i 
\label{y_wm}
\end{equation}

\noindent Based on (\ref{D_i}) and (\ref{y_wm}), \textcolor{black}{ $y_{wm_i}$ can be expressed as}:

\begin{equation}
   y_{wm_i} =
    \begin{cases}
      2^n \cdot y_i + bn_i & \text{if $y_i \in [-T , T]$}\\
      y_i + bn_{max}T + bn_{max} & \text{if $y_i > T$}  
      \\
      y_i - bn_{max}T & \text{$y_i < -T$}  
    \end{cases}
    \label{eq:wmark}
\end{equation}

It should be noted that the carrying measurements occupy the upper bits in $y_{wm_i}$ and encrypted chunks the lower bits. It is a consequence of modifying the original algorithm by taking the average of all measurements instead of that of neighbors.  

\textcolor{black} {By choosing a large enough threshold $T$, all measurements become eligible for insertion. This leads to authorised users not needing a position map in order to perform the extraction.}

Since the algorithm shifts the measurements values in order to expand the prediction error, the dynamic range will expand based on the thresholds and the inserted number of levels.

For a set of patterns \textcolor{black}{ $\phi \in \mathbb{R}^{L\times S}$ } 
of type scrambled Hadamard and a scene to be acquired $x \in \mathbb{R} ^{S\times 1}$ with values $x_i \in [0,1]$, according to eq. (\ref{eq_cs}), the theoretical maximum measurement value occurs when all the elements of 1 in $\phi$ correspond to areas of maximum light in the scene ($x_i = 1$):

\begin{equation}
y_{max} = 2^{-1} \cdot S
\end{equation}

\noindent Similarly, the theoretical minimum would be:

\begin{equation}
y_{min} = - y_{max} = - 2^{-1} \cdot S
\end{equation}

\noindent For an image of $256\times256$ bits, the range is theoretically $[-2^{15},2^{15}]$, which requires a 16 bits representation. \textcolor{black}{By embedding data, even if only using a single insertion level, an overflow appears forcing the  representation on 32 bits.}

\section{The Insertion Capacity}
\label{sec_capacity}

\begin{figure}
    \centering
    \includegraphics[width=0.6\linewidth]{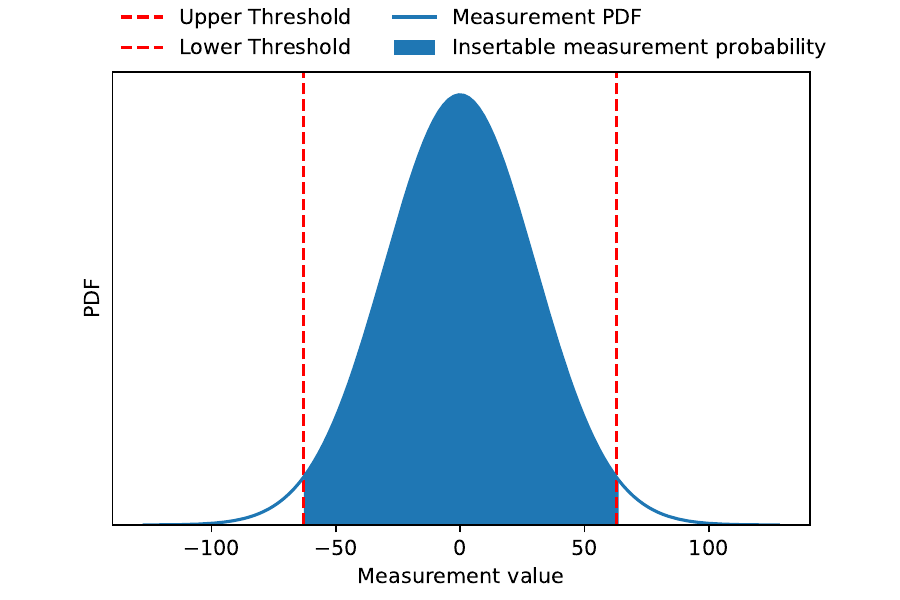}
    \caption{An example of 
 measurement distribution for scrambled Hadamard matrix. The blue area delimited by the thresholds $-T$ and $T$ is the probability of insertable measurements.}
    \label{fig:inssertable-prob}
\end{figure}

The scrambled Hadamard matrix usually behaves like a random Gaussian matrix \cite{Gan2008FastCI}. 
The obtained measurements have a Gaussian distribution with zero mean (Fig. \ref{fig:inssertable-prob}). 
The standard deviation depends on scene sparsity: the more sparse the scene, the lower is the standard deviation.

Knowing the distribution and the thresholds $[-T, T]$, the number of insertable measurements $M$ can be estimated by multiplying the total number of measurements $L$ by the probability $P$ to be between thresholds: 

\begin{equation} 
M = L*P_{T,\sigma} = L*\int_{-T}^{T} \frac{1}{{\sigma \sqrt {2\pi } }}e^{{{ - x ^2 } \mathord{\left/ {\vphantom {{ - x ^2 } {2\sigma ^2 }}} \right. \kern-\nulldelimiterspace} {2\sigma ^2 }}}dx
\label{eq_insertion_probability}
\end{equation}

If the insertion is on  $n$ levels, then the maximum number of bits that could be inserted is  
\textcolor{black} {$L*P_{T,\sigma}*n$.}
In fact, this is only an upper limit since there is always a nonzero probability to use such measurements as data.

Suppose that $x$ is the true number of measurements that, at the end of on-the-fly \textcolor{black}{insertion} process, are carrying data. Then the following equation holds:

\begin{equation}
    x + q*x*\frac{n}{16}=L*P_{T,\sigma}
    \label{eq: trueNo}
\end{equation}

\noindent where \textcolor{black}{$x*n/16$ is the total number of measurements - insertable or not - sacrificed in view of insertion (the rounding was neglected)} and $q$ is the probability to have insertable measurements among them. 
\textcolor{black}{The true number of measurements that carry data, derived from eq. (\ref{eq: trueNo}) is:}

\begin{equation}
    x = \frac{P_{T,\sigma}*L}{1+q*n/16}
    \label{eq:x_wm}
\end{equation}

Hereinafter, we calculate $q$ in the hypothesis that we need two carrying measurements for inserting one, meaning a number of insertion  levels \textcolor{black}{$n$} between $8$ and $16$. Such situation occurs when we have a string of 3 insertable measurements with, eventually, one or more non-insertable measurements between the first and second ones:

\begin{equation}
\begin{split}
    q &=P^3+P^3*(1-P)+P^3*(1-P)^2+...= \\
    &=P^2*[1-(1-P)^m] \approx P^2
\end{split}
\end{equation}

\noindent where $m$ is a the number of terms. The term $(1-P)^m$ becomes quickly negligible, being a probability.
\textcolor{black}{For loose thresholds, $P=1$ and likewise $q=1$. As the thresholds stretch around zero, the probability $P$ reduces, the terms $(1-P)^m$ start to count and the approximation becomes coarser. For $T=3*\sigma$, the error in approximating $q$ is $1\%$ and for $T=2*\sigma$, it is $5\%$. The error estimation was done for $m=1$. For higher $m$, $(1-P)^m$ is lower than $10^{-2}$. We do not encourage the use of small thresholds because it reduces the insertion capacity and fragilizes the method as explained in Section VI.}

With this approximation, the true number of measurements carrying data:

\begin{equation}
    x\approx \frac{P_{T,\sigma}*L}{1+P_{T,\sigma}^2*n/16}
    \label{eq:x_wm}
\end{equation}

The capacity in number of inserted bits is $C=x*n$ and the relative capacity $C_r$ is:

\begin{equation}
    C_r=\frac{C}{L} \approx \frac{P_{T,\sigma}*n}{1+P_{T,\sigma}^2*n/16}
    \label{eq:CrT}
\end{equation}

There are two mechanisms to control the capacity and finally, the distortion: the threshold $T$ through $P_{T, \sigma}$ and the number of insertion levels $n$. The former can provide a fine tuning, the latter a coarse one.

If we  give up to the fine tuning by working with large enough thresholds such to have $P_{T,\sigma}=1$, the relative capacity simplifies to:

\begin{equation}
    C_r \approx \frac{n}{1+n/16}
    \label{eq:Cr}
\end{equation}

\noindent which has the advantage of being independent of image sparsity.

 The fine tuning by controlling $T$ is limited by the constraint of having a representation on 32 bits after data embedding. 
 In this case, the range of $y_{wm}$ should be $[-2^{16},2^{16}]$, meaning:

 \begin{equation}
 y_i + bn_{max}T + bn_{max} < 2^{16}
 \label{Tmax}
 \end{equation}
 
 \begin{equation}
 y_i - bn_{max}T > -2^{16}
 \end{equation}

\noindent From eq. \ref{Tmax}, it follows that the threshold can be at maximum:

\begin{equation}
 T_{max} = \frac{2^{15}}{2^n-1}
 \end{equation}

\noindent For 13 levels for instance, $T_{max}$ is 3 meaning a very low capacity of insertion and a low distortion of the recovered image. It is a mechanism that can be effective only at lower $n$. 

 Figure 4 depicts $C_r$ as a function of measurements standard deviation, parametrized for 8 to 13 insertion levels and the corresponding $T_{max}$. The plots follow a similar trend. The capacity is constant up to a certain standard deviation, then decrease because of insufficient measurements available for carrying data (eq. \ref{eq:CrT}).  
 With dashed line, is the capacity controlled only by the number of insertion levels (eq. \ref{eq:Cr}).

 \begin{figure}[t]
    \centering
 \begin{tabular}{cc}
  \includegraphics[width=0.5\linewidth]{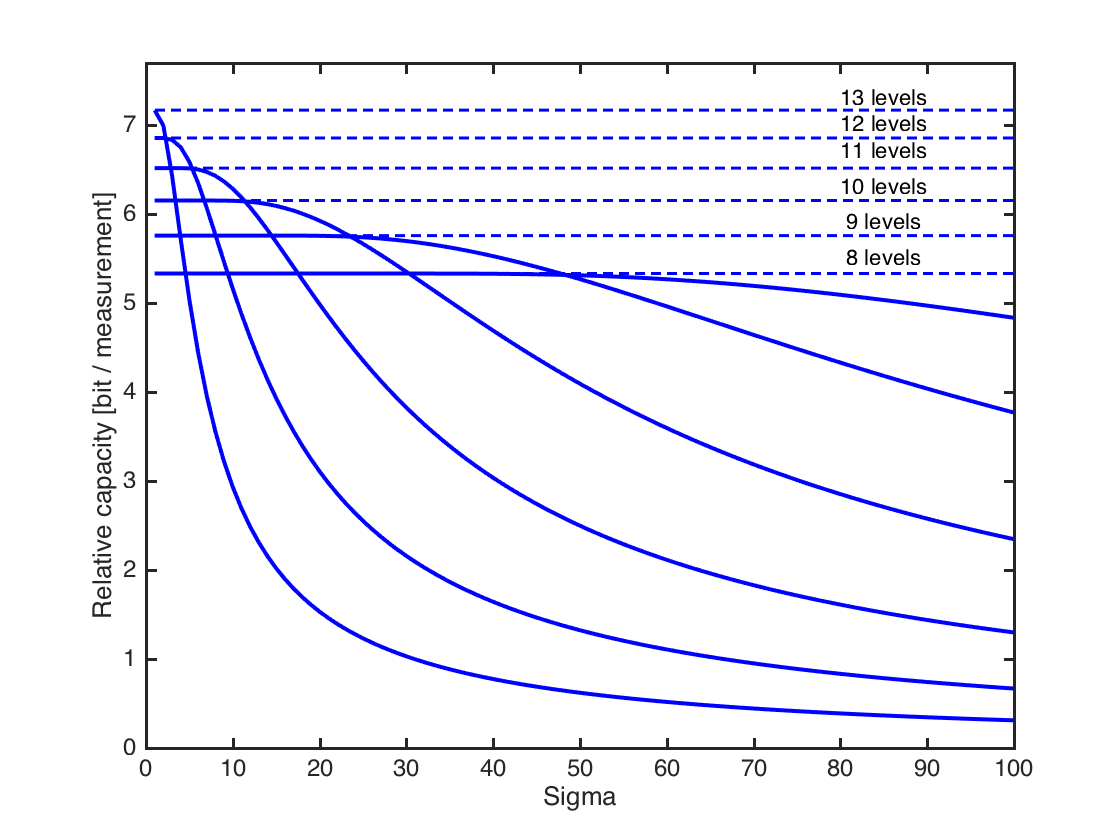}&  \includegraphics[width=0.49\linewidth] {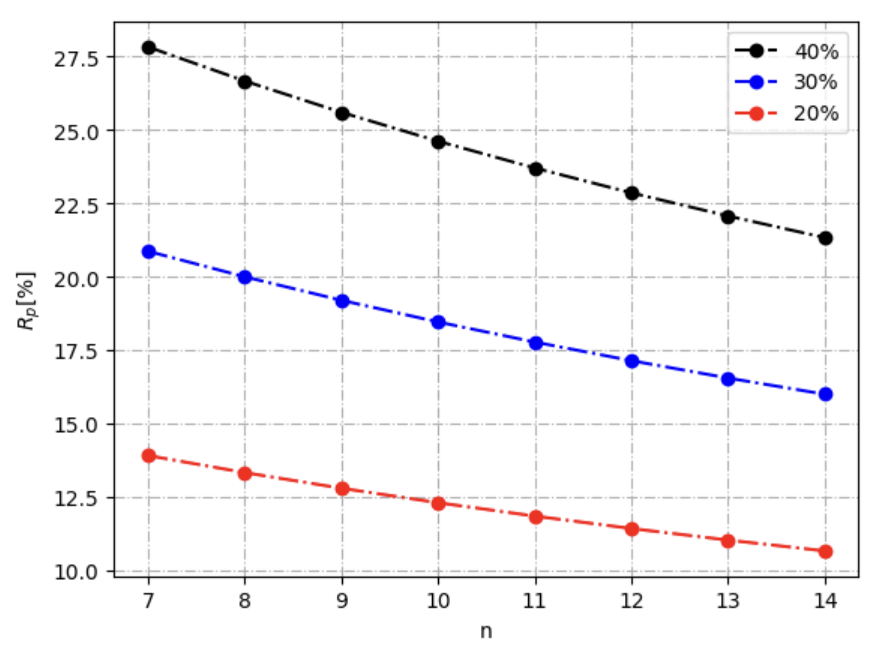}\\

  \makecell{\textbf{Figure 4.} The theoretical relative capacity \\ as a function of measurements standard \\ deviation for $T_{max}$ and parametrized by $n$. \\ With dashed line, the capacity \\ calculated for no restrictive thresholds.} &  \makecell{\textbf{Figure 5.} The percentage of measurements \\after embedding for three initial numbers of \\ measurements: 40\%, 30\% and 20\%.}
 
  \end{tabular}
    \label{fig:figure1}
\end{figure}
\setcounter{figure}{5}

 The number of remaining measurements after insertion and available for a low quality reconstruction is:

  \begin{equation}
    R_p \approx L-\frac{C}{16}=L(1- \frac{C_r}{16})
\end{equation}

 In Fig. 5, $R_p$ is plotted vs. the number of insertion levels for three initial number of measurements given as percentage.

\section{The distortion of the reconstructed image}

In partial encryption, the image reconstructed from the encrypted samples is visibly distorted. In our method, the source of distortion is twofold: less measurements are available for the reconstruction since part of them are removed in view of embedding and the remaining ones are distorted by embedding or shifting.  
These two effects compete for the same goal, which is the image controlled distortion.

\begin{figure}
    \centering
    \includegraphics[width=0.5\linewidth]{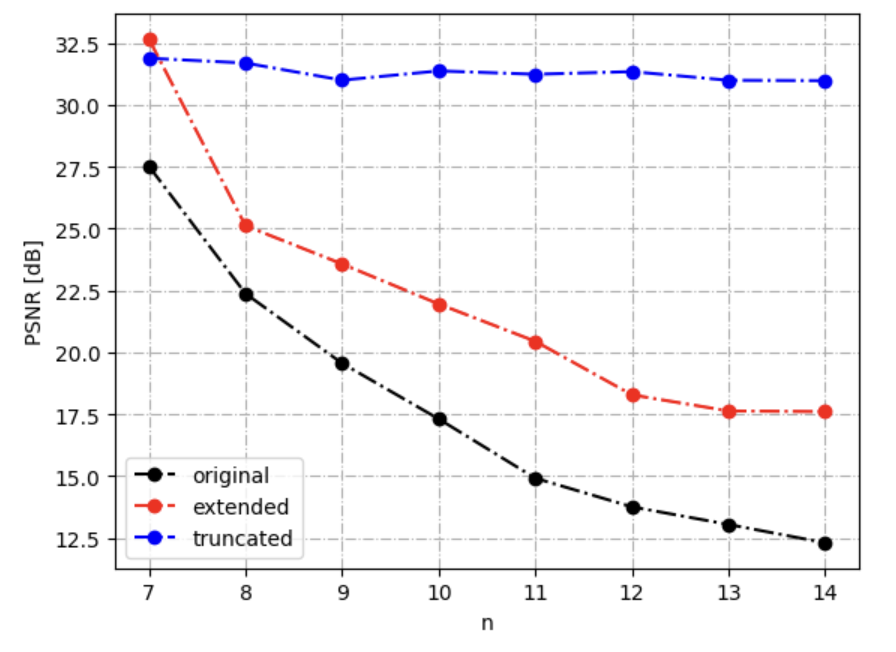}
    \caption{The distortion as a function of insertion levels in three instances: reconstruction from truncated measurements with no embedding (blue), marked measurements extended to contain the measurements to be embedded too (red), and including both embedding and truncation (black).}
    \label{fig:distortion_source}
\end{figure}

To evaluate the contribution of the two sources, we have reconstructed images in two instances: by inserting data and meanwhile, keeping all the measurements and by removing $M_{embed}$ measurements without any embedding. Figure \ref{fig:distortion_source} depicts the image distortion given as Peak Signal to Noise Ratio (PSNR)  versus the number of embedded levels, in these two instances. 
The plots are obtained experimentally by averaging the PSNR over 359 patches from the sky image in section \ref{Experiments}. Loose thresholds were used, so that all measurements are considered eligible for embedding.

For this kind of image that is rather sparse, at 40\% initial measurements, the main source of distortion is the data insertion or shifting, which alters the values in $y_{wm}$. In comparison, the reconstruction quality when truncating the measurement sequence does not present major variations when depicted as a function of insertion levels. To be sensitive to truncation, $L$ should have been lower.

If tight thresholds are used, the number of measurements selected for embedding is reduced. To have a fully contribution of the truncation mechanism, loose thresholds - enough to not matter - should be used.
In this case, the distortion  can be controlled only through the number of insertion levels.

\section{\textcolor{black}{Data Integrity Aspects}}

The partial encryption is prone to ECA attack that consists in estimating the encrypted coefficients based on some a priori knowledge like, for instance, the position of low frequency coefficients in DCT based image compression.

In CS, due to the random projection, there are no privileged measurements. 
In our method, the weakness is given by the fact that the encrypted measurements  are placed \textcolor{black}{at insertion,}
 
 on the less significant bits of the carriers. 

\begin{figure}[t]
    \centering
 \begin{tabular}{cc}
  
  \includegraphics[width=0.3\linewidth]{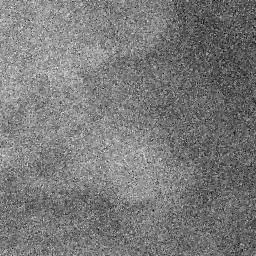}&  \includegraphics[width=0.3\linewidth] {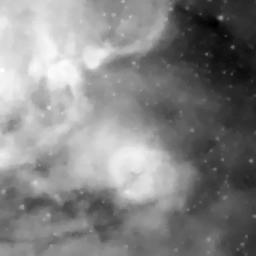}\\
   
  a) PSNR 14.67 dB, L=$40\%$ & b) PSNR 31.03 dB, L=$40\%$ \\ 
    
  \includegraphics[width=0.3\linewidth]{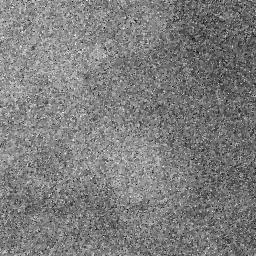} &  \includegraphics[width=0.3\linewidth] {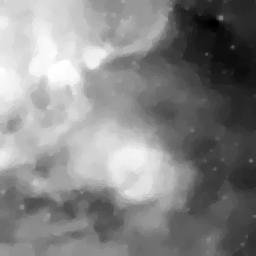}\\
  c) PSNR 13.30 dB, L=$18\%$ & d) PSNR 30.12 dB, L=$18\%$
  \end{tabular}
\caption{Example of ECA attack: (a,c) Patch reconstructed with \textcolor{black}{insertion on 10 levels}, (b,d) The same patch recontructed with clean measurements and reduced Hadamard matrice (after EDA attack).}
\label{fig:EDA}
\end{figure}

If loose thresholds are used, by several trials the encrypted chunks can be discarded and a better quality image can be reconstructed with the reduced Hadamard matrix. 
An example is given in Fig. \ref{fig:EDA}. After ECA, the PSNR is improved  by 16.36 dB when the initial percentage of measuremets is $40\%$ and by 16.82 dB for 18\% initial measurements. In this second case, the truncation mechanism becomes effective. 
The PSNRs are calculated by respect with the reconstruction with full number of measurements.

\textcolor{black}{Even with this remarkable improvement in PSNR, the recovered measurements are not complete. The embedded measurements that were discarded by ECA, are missing. According to Fig. 5, after ECA,  the reconstruction is  done with less than 25\% measurements instead of 40\% in the first case and less than 12.5\% instead of 18\% in the second one. The embedded measurements can be recovered only if the user has access to the secret key used to protect them. The PSNR, which is 32.03 dB and 30.12 dB respectively, confirms missing data.}

\textcolor{black}{Data integrity protection is obtained at the expense of volume increase. 
The number of measurements sacrificed for embedding does not compensate for extending data representation from 16 to 32 bits. The compression rate calculated in Section VII shows how this overflow depends on the number of insertion levels.}

If tight thresholds are used, it is more difficult to recover clean measurements because of $b_{nmax}T$ in eq. (\ref{eq:wmark}) that needs the knowledge of $T$ in order to be calculated.

\begin{figure}[t]
    \centering
    \includegraphics[width=0.7\linewidth]{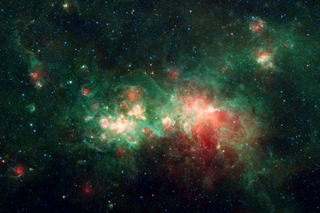}
    \caption{The image of the W51 nebula obtained by NASA's Spitzer Space Telescope.\\  The image was used to generate the simulated data \cite{Spitzer2020}.}
    \label{fig:ssc2020-14a}
\end{figure}

\begin{figure}[h]
    \centering
    \includegraphics[width=.3\linewidth]{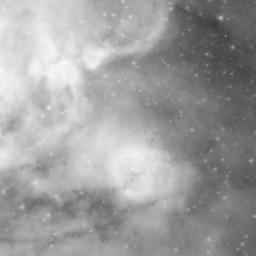}
    \includegraphics[width=.3\linewidth]{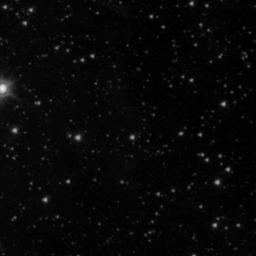}
    \caption{Two examples of patches used for tests: a dense one (left) and a sparse one (right).}
    \label{fig:example_slices}
\end{figure}

\section{Experimental Results}
\label{Experiments}

We experimentally test our method under the following aspects: embedding capacity, distortion, and impact on data volume. 

\textcolor{black}{The experiments are carried out on simulated and real data. The simulated data are obtained from an image of the W51 nebula (Fig. \ref{fig:ssc2020-14a}) converted to grayscale using the weighted method, normalized to have values between 0 and 1 and split into 359 patches of 256 x 256 pixels each.
Fig. \ref{fig:example_slices} shows two patches with different sparsities.} 
\textcolor{black}{We simulated the acquisition by calculating random projections on a Hadamard matrix generated using Sylvester’s construction.
The matrix only contains 1 and -1 elements. The elements in the first row are all positive, while all the other rows have an
equal number of positive and negative elements. The first row was excluded since it leads to a first measurement with a double maximum value compared
to the others. The matrix columns are randomly scrambled before image projection \cite{Gan2008FastCI}. We use 40\% measurements meaning that 60\% of the matrix rows are discarded. To reconstruct the images from the embedded measurements, the matrix size is reduced accordingly.}

\textcolor{black}{
The real data consist of a set of measurements acquired with our setup in \cite{Petrovici2016Single}. The imaged object is a 0.5 mm thick
sheet of metal with machined patterns in it, and with a total dimension of 13mm×13mm (Fig. \ref{fig:metalSheet}). The object was exposed to homogenized light from a halogen lamp. }
\textcolor{black}{Since the construction of the SLM in Single Pixel camera
only permits the use of sensing matrices composed by 0 or 1,
the real data were measured with a S-matrix having randomly scrambled columns. Such a matrix ca obtained from the Hadamard matrix by removing its first row and column and turning all the negative values to 0 \cite{Sloane1979Multiplexing}, followed by randomly permuting its columns.}

\textcolor{black}{For image reconstruction, either from simulated or real data, we use an implementation of the fast iterative shrinkage-thresholding algorithm (FISTA) presented in \cite{Beck2009Fast}.}

For embedding data, we considered a loose threshold of $T=500$, which ensures that all measurements are eligible for insertion.

When analyzing the results, we must take into consideration that an unauthorized user should still be able to reconstruct a lower quality version of the original image \textcolor{black}{using the modified measurements}.

\begin{figure}
    \centering
    \includegraphics[width=0.5\linewidth]{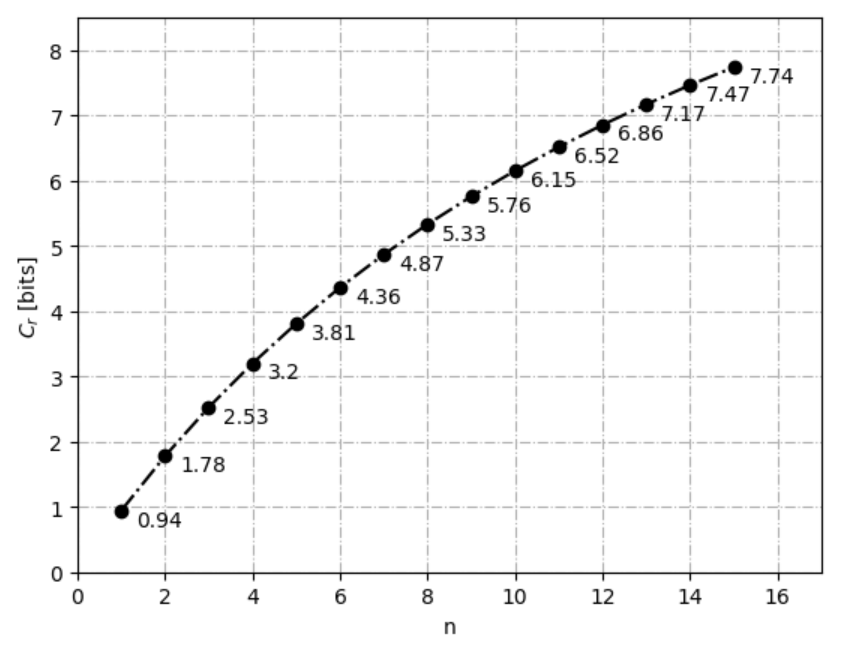}
    \caption{The relative capacity $C_r$ vs. the number of insertion levels $n$, for loose thresholds \textcolor{black}{($T=500$).}}
     
    \label{fig:cr_no_insertion_levels}
\end{figure}

Since in our experiment all measurements are eligible for insertion, the capacity can be calculated directly using eq. (\ref{eq:Cr}).
In Fig. \ref{fig:cr_no_insertion_levels}, the relative capacity $C_r$ is plotted against the number of insertion levels $n$. The curve has an ascending trend, with capacity values ranging from $0.9$ bits/measurement for 1 level to $7.5$ bits/measurements for 14 levels.

To evaluate the \textcolor{black}{image distortion obtained by embedding,}
 
we calculate the PSNR taking as reference the image reconstructed from the original measurements.
Figure 11 plots the median of the PSNR of all 359 patches \textcolor{black}{for a loose threshold ($T=500$) and a more restrictive one ($T=50$).}
A drop in the reconstruction quality can be noticed after 6 levels. 
Up to 7 insertion levels, the distortion is not visible.   
For this reason, we restrict our analysis at $n\in [7,14]$.
\textcolor{black}{We repeated the experiments for a tight threshold $T=50$. The plot in Fig. 11  shows a significantly higher distortion for small $n$ but results close to loose thresholds for $n>8$.}

\begin{figure}[t]
    \centering
 \begin{tabular}{cc}
  \includegraphics[width=0.49\linewidth]{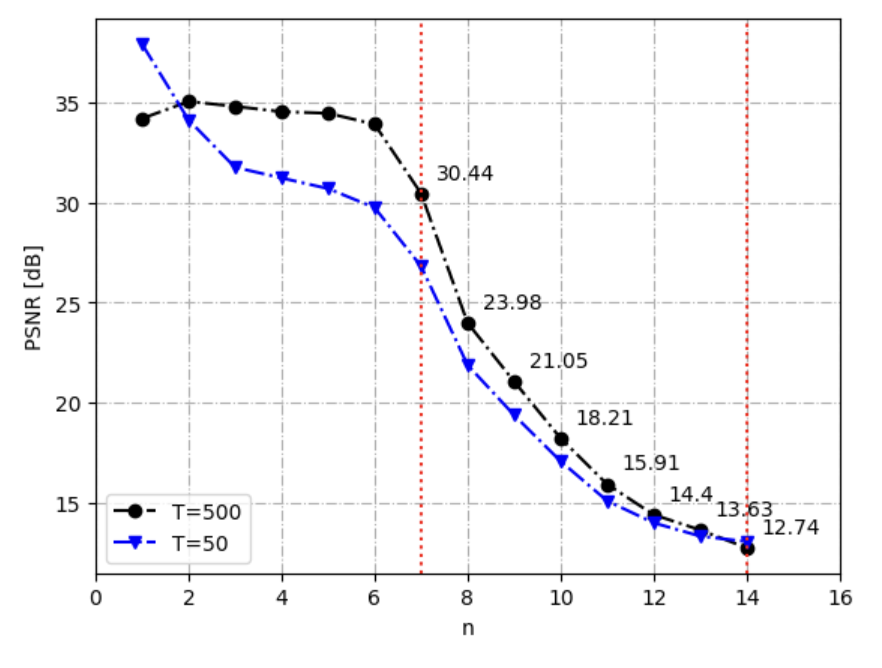}&  
  \includegraphics[width=0.5\linewidth]{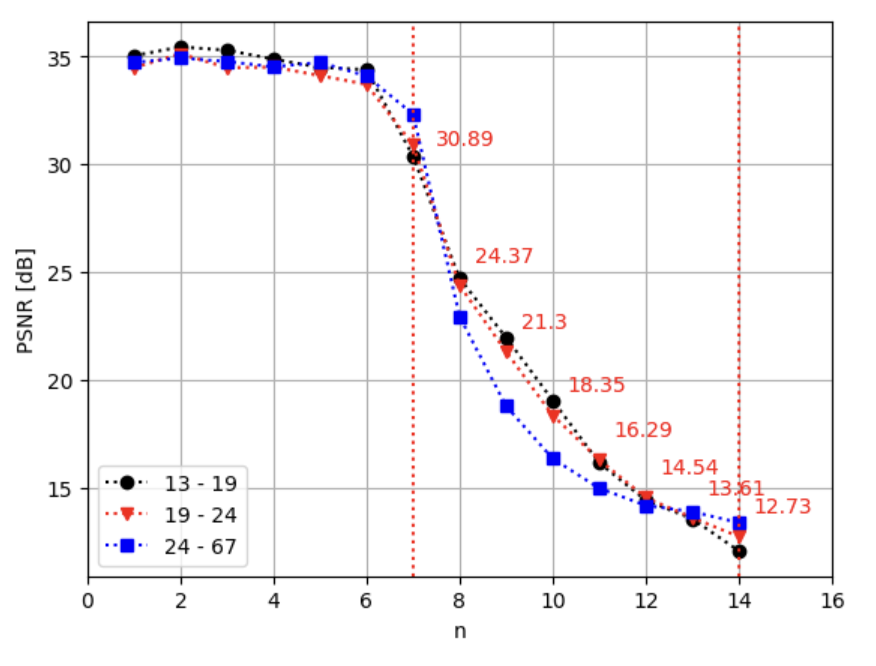}\\

  \makecell{\textbf{Figure 11.} The median of PSNR  \\  experimental values  against $n$ for \\  $T=500$ and $T=50$. \\ 
    The vertical dotted lines marks the range \\  of interest for the application.} &  \makecell{\textbf{Figure 12.} The median of PSNR  \\ experimental values for 3 ranges of  \\ image standard deviation and  \\  loose thresholds.}
 
  \end{tabular}
    \label{fig:figure1}
\end{figure}
\setcounter{figure}{12}

\textcolor{black}{We deepened the analysis and we split the image set into 3 subsets corresponding to the following 3 ranges of pixels standard deviation: [13,19], [19,24] and [24,67]. The ranges were chosen such to cover the standard deviations of all images and to have approximately the same number of images in each subset.  In natural images, the standard deviation gives an indication about the image sparsity, which is the main parameter in the CS reconstruction from a fixed percentage of measurements. The plots of PSNR vs. $n$ are quite similar for the 3 ranges (Fig. 12). }

\begin{figure}[!h]
    \centering
 \begin{tabular}{cc}
  
  \includegraphics[width=0.3\linewidth]{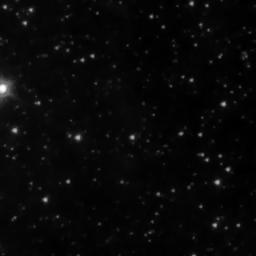}&  \includegraphics[width=0.3\linewidth] {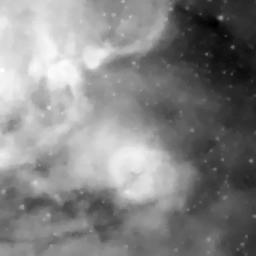}\\
   
  a) Original & e) Original\\ 
   
  \includegraphics[width=0.3\linewidth]{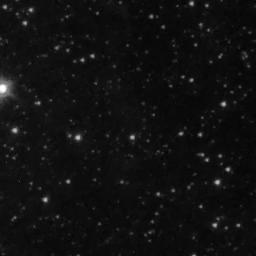} &  \includegraphics[width=0.3\linewidth] {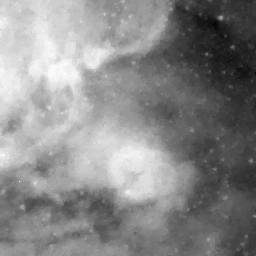}\\
  b) PSNR 29.18 dB & f) PSNR 32.23 \\

  \includegraphics[width=0.3\linewidth]{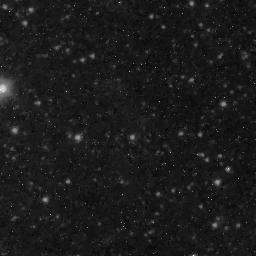}&  \includegraphics[width=0.3\linewidth] {dense_10_bits.jpeg}\\
   
  c) PSNR 21.89 dB & g) PSNR 14.67 dB \\ 
   
  \includegraphics[width=0.3\linewidth]{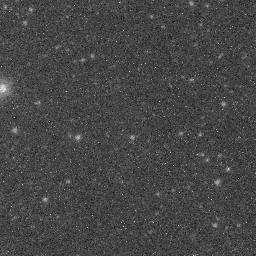} &  \includegraphics[width=0.3\linewidth] {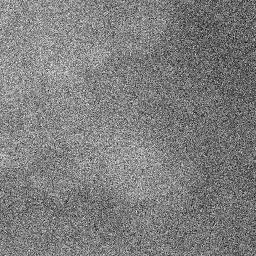}\\
  d) PSNR 12.06 dB & h) PSNR 13.45 dB
  \end{tabular}
    \caption{Reconstruction examples for a sparse and a dense patch: (a,e) originals obtained with the 40\% genuine measurements; (b,f) images after performing insertion on 7 levels;  (c,g) images after insertion on 10 levels;  (d,h) images after insertion on 14 levels.}
    \label{fig:example_slices_sparse_dense}
\end{figure}

Examples of image degradation are shown in Fig. \ref{fig:example_slices_sparse_dense}. 
For $7$ insertion levels the quality of the two patches is rather good.
At $10$ insertion levels, the distortion is quite strong, especially for the dense patch that has a PSNR of $14.67$ dB.
Its content becomes almost undiscernable at 14 levels.

\textcolor{black}{The PSNR median at 10 levels is $18.21$ dB for the whole set (Fig. 11). On the 3 ranges considered for the standard deviation, it is $19.05$ dB, $18.35$ dB and $16.45$ dB, respectively. Given the images visual quality and the small variations of the PSNR with the standard deviation range, we adopted 10 as being the appropriate number of insertion levels.}

\begin{figure} [t]
    \centering
    \includegraphics[width=0.5\linewidth]{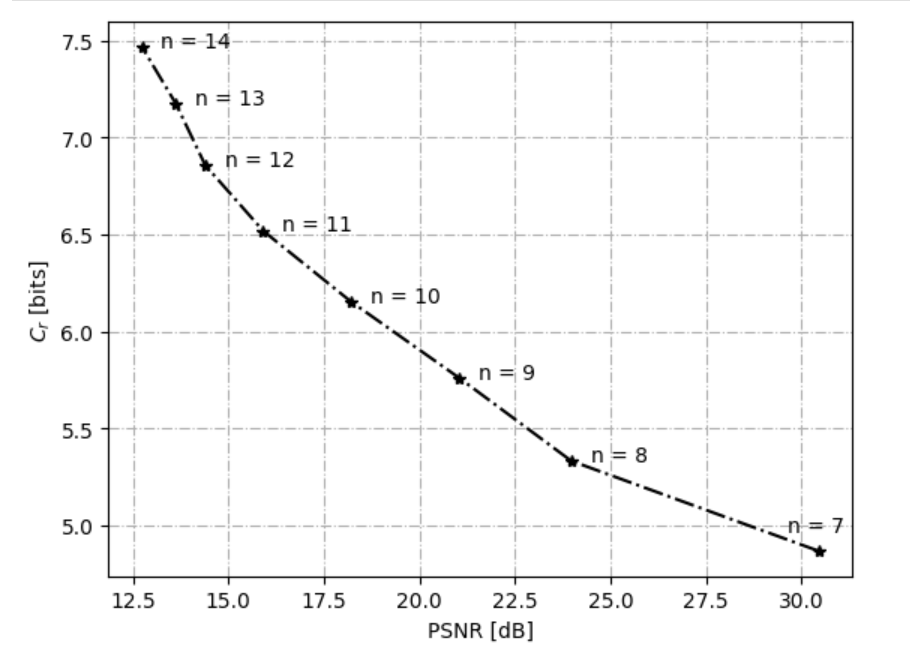}
    \caption{Relative Capacity $C_r$ as a function of PSNRs median.}
    \label{fig:CR(PSNR)}
\end{figure}

The distortion and the capacity are closely related. Figure \ref{fig:CR(PSNR)} shows the correlation between the relative capacity $C_r$ and the PSNR. 
As expected, both capacity and distortion increase with the number of insertion levels.

The data insertion by our algorithm modifies the data volume. There are two opposite effects that compete: sacrificing measurements for insertion reduces the data volume and data embedding that extends the data representation from 16 to 32 bits.
These effects are summarized in the compression rate equation below:

\begin{equation}
    r = 2 * x / L
\end{equation}

\noindent For loose thresholds, the number of measurements $x$ in eq. (\ref{eq:x_wm}) becomes:

\begin{equation}
    x \approx \frac{L}{1+n/16}
\end{equation}

\noindent which leads to the following equation for the compression rate:

\begin{equation}
    r \approx \frac{2}{1 + n/16}
\label{eq:eq_compression_rate}
\end{equation}

\noindent The compression rate shows that the algorithm always expands the data volume for a number of insertion levels less than $16$. 
Also, it should be noted that given the loose thresholds, the data volume is not influenced by either the image size or the content. Only the number of insertion levels matters. 
Figure 15 depicts the evolution of $r$ for the analyzed range of insertion levels.  
\textcolor{black}{The compression rate is always higher than 1, showing a data overflow varying from 39\% for $n=7$ to 7\% for $n=14$. }

\textcolor{black}{The data throughput of SPCs is limited by the Digital Micromirror Device (DMD). Texas Instruments DMD products have binary pattern rates ranging from 2,500 Hz to 32,000Hz \cite{TI}. By multiplying these rates by 32 bits, which is the measurement binary representation after data hiding, a maximum throughput of about 1 MHz is to be expected. The volume expansion by only 7-39\% takes into consideration the reduction of the number of measurements to be transmitted. This means a shorter transmission time but here, we must take into consideration also the processing time necessary to data hiding. }

To jointly evaluate the data expansion and image distortion, we plotted in Fig. 16 the rate-distortion curve. As the number of insertion levels increases, the distortion becomes more evident and the data expansion  less important. 

For $10$ levels, a number that we considered appropriate for our application, the measurements volume increases by $23\%$.
With respect to image size and taking into account that the sampling rate is $40\%$, it follows the data volume after insertion represents $2*0.4*1.23$ from the image size, which means an image compression rate of $98\%$. CS expands data representation by $2$, from $8$ bits/pixel to $16$ bits/measurement.

\begin{figure} [t]
    \centering
 \begin{tabular}{cc}
  \includegraphics[width=0.5\linewidth]{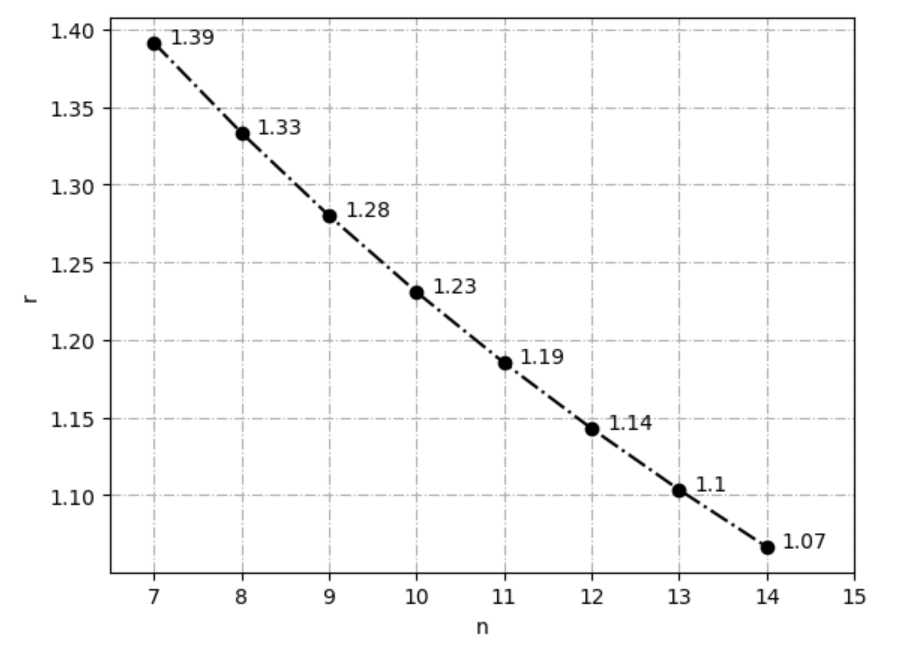}&  \includegraphics[width=0.49\linewidth] {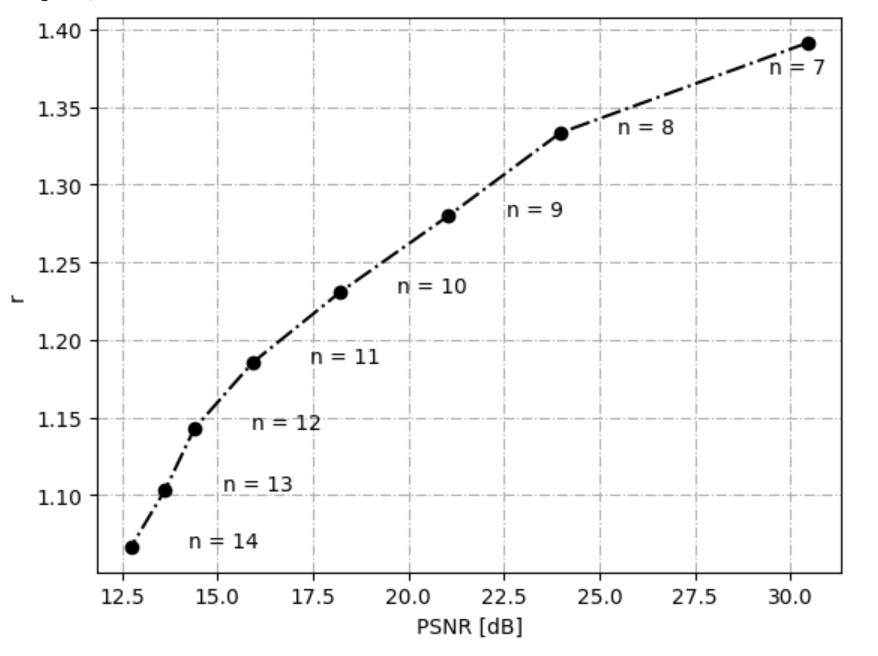}\\

  \makecell{\textbf{Figure 15.} The compression rate vs. $n$ \\ for insertion \\ with loose thresholds.} &  \makecell{\textbf{Figure 16.} The rate-distortion curve for \\ insertion with loose thresholds.}
 
  \end{tabular}
    \label{fig:figure1}
\end{figure} 

\setcounter{figure}{16}

\textcolor{black}{We measured the time of execution and necessary memory for both data hiding and data extraction for $256\times256$ pixels images, 40\% measurements and loose thresholds (Table I). The simulations were performed on macOS 13.4 with 3.5 GHz CPU and 8 GB RAM, using the software Python 3.9.6.
It is to note that the time and memory increase with $n$ and that they are higher for data extraction. Table I also includes the values obtained when inserting on 10 bits and using $L = 20\%$, respectively $L = 30\%$. As expected, they are lower for both execution time and memory requirements when compared to the $L = 40\%$ and $n=10$ values.}

\begin{table*}[b]
\centering
\caption{Execution time and required memory for $L=40\%$ and loose thresholds. }
\vspace{2mm}
\begin{tabular}[ht]{|c|c|c|c|c|}
\hline
\makecell{Insertion levels\\$n$} & 
\makecell{Insertion time\\$[ms]$} & \makecell{Extraction time\\$[ms]$} & \makecell{Insertion \\ memory$[MB]$} &  \makecell{Extraction \\ memory$[MB]$}\\
\hline
 
7 & 179.06 &  1159.12 &  1.39 &  2.85  \\ 
 \hline
 8 & 186.24 & 1259.10  & 1.45 & 3.01  \\ 
\hline
9 & 213.02 & 1359.29  & 1.48 & 3.20 \\ 
\hline
10 & 231.15 & 1447.98  & 1.54 & 3.39 \\ 
\hline
11 & 247.34 & 1537.39 & 1.59 & 3.51 \\ 
\hline
12 & 256.72 & 1621.28  & 1.62 & 3.64 \\ 
\hline
13 & 273.28 & 1679.53  & 1.65 & 3.82 \\ 
\hline
14 & 281.57 & 1748.07  & 1.70 & 3.89 \\ 
\hline
$n=10, L=20\%$ & 121.25 & 778.05  & 0.78 & 1.71 \\ 
\hline
$n=10, L=30\%$ & 186.71 & 1181.71  & 1.20 & 2.51 \\ 
\hline

\end{tabular}
\label{tab:1Hdeblurring}
\end{table*}

\textcolor{black}{The experiments on smaller patches of 128x128 pixels have shown on average, a higher PSNR. If the user chooses to work with patches sizes other than $256\times256$ pixels, the distortion can be tuned from the number of insertion levels.}

\textcolor{black}{We compare our method with \cite{Wang2019}, which is close to ours as goal and approach. Similar to us, the authors of \cite{Wang2019} perform a partial encryption of CS measurements and hide information. Unlike our method, the encryption concerns the sign of part of measurements and the hidden data is side information. The tuning of distortion is obtained via the percentage of encrypted measurements. For $256\times256$ images and 50\% measurements, at an encryption percentage of $0.62\%$ the distortion of the reconstructed image is $16.7$ dB, similar to our result. The data hiding in \cite{Wang2019} is done by histogram shifting, which supposes an estimation of the measurements distribution. Our method has the advantage of being blind and that of an on-the-fly insertion.}

\textcolor{black}{Two other methods related to our work are \cite{Xiao2020},\cite{Thakkar2017}. Although the CS does not concern scene acquisition, it is used to pre-process the cover or watermak images. In \cite{Xiao2020}, for pre-processing, the authors use prediction coding based on CS progressive recovery and encryption by XOR with a random stream generated with a secret key. Side information is inserted into the encrypted data. The resulting insertion capacity is up to 4 bpp for a recovered image quality higher than 39 dB. In \cite{Thakkar2017}, the CS is applied on the Principal Components (PCs) of watermark image to get CS measurements that are further embedded into the HL subband of the wavelet-transformed cover image. The method has three security layers, one from PCs and two from CS. The method reaches insertion rates of around 2 bpp and a PSNR above 30 dB for both watermark and cover image.}

\textcolor{black}{The insertion in real data with our method led to lower PSNRs compared to simulated measurements. At 10 insertion levels, the PSNR is $10.48$ dB for the real image and about $14$ dB for the images reconstructed from simulated measurements. The real images are shown in Fig. \ref{fig:metalSheet}. In all three cases i.e., 7, 10 and 14 insertion levels, the PSNRs are under those obtained for synthetic data.} 

\begin{figure}[!h]
    \centering
 \begin{tabular}{cc}
  
  \includegraphics[width=0.3\linewidth]{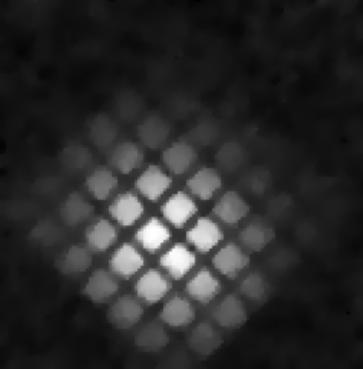}&  \includegraphics[width=0.3\linewidth] {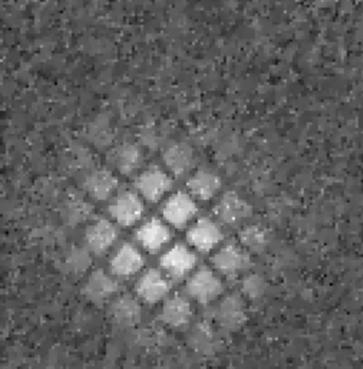}\\
   
  a) Original & b) PSNR 12.69 dB\\ 
   
  \includegraphics[width=0.3\linewidth]{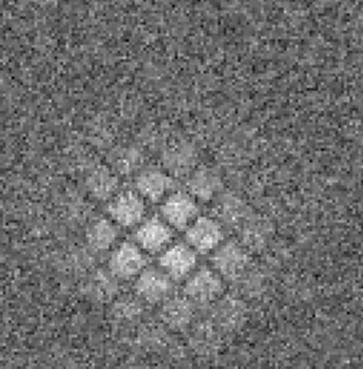} &  \includegraphics[width=0.3\linewidth] {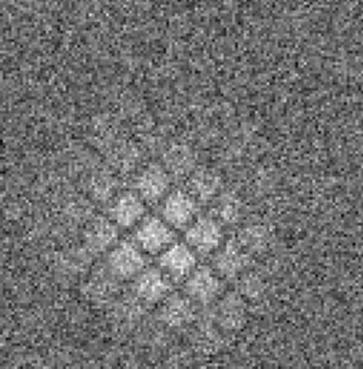}\\
  c) PSNR 10.48 dB & d) PSNR 9.94 dB \\
 
  \end{tabular}
\caption{Images reconstructed from real measurements: (a) Original obtained with 40\% measurements; (b) image after data insertion on 7 levels;  (c) image after insertion on 10 levels;  (d) image after insertion on 14 levels. The PSNR is calculated by respect to the Original.}
    \label{fig:metalSheet}
\end{figure}

\section{Conclusion}

The method we propose for securing CS streams is effective for multiple insertion levels. For the synthetic data used in our experiments, the appropriate number of levels has been 10, with an average distortion of 18 dB, which lets the image content still discernible for an unauthorized user. For the real data and smaller images, the distortion was higher, suggesting that a smaller number of insertion could be used. However, since the insertion capacity and the compression rate are directly impacted, we recommend to use, as much as possible, a high number of insertion levels. 

 The control of image distortion is done mainly through the number of insertion levels. If an extension beyond 32 bits is tolerated for the modified measurements, a second mechanism of control i.e., the insertion threshold $T$ can be used. It adds a supplementary security layer.

The method expands the measurements volume by $40\%$ to $6\%$ for $n\in[7,14]$. At 10 levels, the volume increases by $23\%$.

The method is sensitive to ECA attacks, especially when the images are very sparse and good reconstructions can be obtained from fewer measurements. However, the number of recovered measurements remains incomplete, and the original image cannot be obtained without the secret key used for encryption.

\section*{Author Contributions}
C.E.Popa: conceptualization, methodology, writing (Section 1, Section 3-7). C.C. Damian: methodology, review. D. Coltuc: conceptualization, writing (Section 2 and 8), review and editing. All authors have read and agreed to the published version of the manuscript.

\section*{Funding}
This research received no external funding.

\section*{Institutional Review Board Statement}
Not applicable.

\section*{Informed Consent Statement}
Not applicable.

\section*{Data Availability Statement}
Data is contained within the article.

\section*{Conflicts of Interest}
The authors declare no conflict of interest.

\end{document}